\documentclass[prl,nofootinbib,superscriptaddress,twocolumn]{revtex4-1}
\usepackage[a4paper,left=1.5cm,right=1.5cm,top=3cm,bottom=3cm]{geometry}

\usepackage{verbatim}
\usepackage{color}
\usepackage{amsfonts,amssymb,mathrsfs,amsmath}
\usepackage{framed}
\usepackage{mdframed}
\usepackage{latexsym}
\usepackage{graphicx}
\usepackage{datetime}
\newdateformat{mydate}{\THEDAY{ }\monthname[\THEMONTH]{ }\THEYEAR}

\allowdisplaybreaks

\usepackage{tikz}
\usepackage{color}
\usepackage{framed}
\usepackage{hyperref}
\hypersetup{colorlinks, citecolor=bluscuro, linkcolor=black, urlcolor=bluscuro}
\definecolor{rossos}{cmyk}{0,1,1,0.55}
\definecolor{bluscuro}{rgb}{0.15, 0.2, .85}
\definecolor{bluchiaro}{cmyk}{1,.3,0.,0.1}

\newcommand{\be}{\begin{equation}}
\newcommand{\ee}{\end{equation}}
\renewcommand{\d}{{\rm d}}

\def\PBH{\text{\tiny PBH}}

\newcommand{\lp}{\left (}
\newcommand{\rp}{\right )}

\def\vk{{\vec{k}}}

\newcommand{\Ic}{\mathcal{I}_c}
\newcommand{\Is}{\mathcal{I}_s}

\newcommand{\Pz}{\mathcal P_\zeta}

\newcommand{\FF}[1]{\widetilde{#1}}

\newcommand{\dd}{{\rm d}}
\newcommand{\OGW}{\Omega_\text{\tiny GW}}

\def\lsim{\mathrel{\rlap{\lower4pt\hbox{\hskip0.5pt$\sim$}}
    \raise1pt\hbox{$<$}}}         
\def\gsim{\mathrel{\rlap{\lower4pt\hbox{\hskip0.5pt$\sim$}}
    \raise1pt\hbox{$>$}}}         

\newcommand{\arXiv}[2]{\href{http://arxiv.org/pdf/#1}{{\tt [#2/#1]}}}
\newcommand{\arXivold}[1]{\href{http://arxiv.org/pdf/#1}{{\tt [#1]}}}

\begin{document}

\title{NANOGrav Hints to Primordial Black Holes as Dark Matter}

\author{V. De Luca}
\email{Valerio.DeLuca@unige.ch}
\address{D\'epartement de Physique Th\'eorique and Centre for Astroparticle Physics (CAP), Universit\'e de Gen\`eve, 24 quai E. Ansermet, CH-1211 Geneva, Switzerland}

\author{G. Franciolini}
\email{Gabriele.Franciolini@unige.ch}
\address{D\'epartement de Physique Th\'eorique and Centre for Astroparticle Physics (CAP), Universit\'e de Gen\`eve, 24 quai E. Ansermet, CH-1211 Geneva, Switzerland}

\author{A.~Riotto}
\email{Antonio.Riotto@unige.ch}
\address{D\'epartement de Physique Th\'eorique and Centre for Astroparticle Physics (CAP), Universit\'e de Gen\`eve, 24 quai E. Ansermet, CH-1211 Geneva, Switzerland}

\address{INFN, Sezione di Roma, Piazzale Aldo Moro 2, 00185, Roma, Italy}

\date{\today}

\begin{abstract}
\noindent
The  NANOGrav Collaboration has recently
published a  strong evidence for a stochastic common-spectrum process that may be interpreted as a  stochastic gravitational
wave background. We show that such a signal can be explained by second-order gravitational waves produced during the formation of primordial black holes from the collapse of sizeable
scalar perturbations generated during inflation. This possibility  has two predictions:  {\it i)} the primordial black holes may comprise the totality of the dark matter with the dominant contribution to their mass function falling  in the range $(10^{-15}\div 10^{-11}) M_\odot$  and {\it ii)} the gravitational wave stochastic 
background will be seen as well by the LISA experiment.

\end{abstract}

\maketitle

\paragraph{Introduction.}
\noindent
The 
NANOGrav Collaboration has recently published an  analysis of 12.5 yrs of pulsar timing data \cite{Arzoumanian:2020vkk} reporting a strong evidence for a
stochastic common-spectrum process. The latter may be compatible with a Gravitational Wave (GW)  signal with  strain amplitude $\sim 10^{-15}$ 
at a frequency $f\sim 3\cdot 10^{-8}$ Hz with an almost  flat GW spectrum, $\OGW(f)\sim f^{(-1.5\div 0.5)}$ at 1$\sigma$-level. In particular, their analysis shows 
  the presence of a stochastic process
across fourty-five pulsars which can be interpreted  in terms of a common-spectrum process strongly preferred against independent
red-noise signals. Despite being in partial  contrast with some other bounds on the stochastic background of GWs, the NANOGrav Collaboration stresses  that the detected signal    is due to an  improved treatment of the intrinsic pulsar red
noise. On the other side, it is important to stress that the NANOGrav Collaboration does not claim a detection
of GWs since the signal does not possess   quadrupole correlations.

The goal of this paper is to show that the NANOGrav signal, if interpreted as a GW background, can be naturally explained by a flat spectrum of GWs inevitably generated at second-order in perturbation theory during the  
formation of Primordial Black Holes (PBHs) in the case in which the latter form from the collapse of large curvature perturbation generated during inflation upon horizon re-entry  (see also Refs.~\cite{revPBH, Green:2020jor} for reviews on PBHs physics and Ref.~\cite{Carr:2020gox} for a review on the constraints on their abundance).

Two nice by-products of this explanation are that {\it i)} the dominant contribution to the PBH mass function falls in the range $(10^{-15}\div 10^{-11}) M_\odot$ where the PBHs can comprise the totality of the dark matter in the universe; {\it ii)} the GW stochastic spectrum propagates to frequencies testable by future experiments, such as LISA \cite{Audley:2017drz}. 

\vskip 0.3cm
\noindent

\paragraph{The PBH abundance.}
The most common formation scenario for PBHs is through an enhancement of the power spectrum of the comoving curvature perturbation $\zeta$ during inflation, at scales much smaller than those probed by CMB observations~\cite{s1,s2,s3}.
During the radiation-dominated phase,  an overdense region collapses to form a PBH at horizon re-entry if the volume-averaged density contrast is larger than a critical value $\delta_c$, which has been found with dedicated relativistic numerical simulations in Refs.~\cite{Harada:2015yda,musco, Germani:2018jgr}. 

We define the comoving curvature perturbation power spectrum as
\begin{equation}
\big<\zeta(\vk_1)\zeta(\vk_2)\big>' =\frac{2\pi^2}{k_1^3} \Pz(k_1),
\label{eq: def P zeta}
\end{equation}
where  we have adopted  the standard prime notation indicating that  we do not explicitly write down the $(2\pi)^3$ times the Dirac delta of momentum conservation.
In comoving slices, the overdensity is usually expressed in terms of the curvature perturbation through the non-linear relation \cite{Harada:2015yda}
  \be
  \label{nonlinear}
 \delta(\vec x)=-\frac{8}{9a^2H^2}e^{-5\zeta(\vec x)/2}\nabla^2e^{\zeta(\vec x)/2},
 \ee
where $a$ is the scale factor and $H=\dot a/a$ the Hubble rate.

Assuming Gaussian curvature perturbations, one can estimate the mass fraction of the Universe that collapses to form PBHs 
at formation by computing the probability $P (\delta)$ that the overdensity is larger than the critical threshold  following 
the Press-Schechter formalism as \cite{revPBH}
\be
\beta (M_\text{\tiny PBH}) = \int_{\delta_c}^\infty \d \delta\,  \frac{M_\text{\tiny PBH}}{M_H} P (\delta),
\ee 
where one has to keep into account the scaling law relating the PBH mass to the horizon mass $M_H$ for overdensities close to the critical threshold for collapse  as \cite{Choptuik:1992jv, Evans:1994pj, Niemeyer:1997mt}
\be
M_\text{\tiny PBH} = \kappa M_H (\delta - \delta_c)^{\gamma_c}
\ee
in terms  of the constants $\kappa = 3.3$ and $\gamma_c = 0.36$ in a radiation-dominated universe \cite{koike,Musco:2004ak,Musco:2008hv,Musco:2012au,Kalaja:2019uju,Escriva:2019nsa}.
The
horizon mass $M_H$ is related to the characteristic comoving frequency of the perturbation as
\be
M_H \simeq 33 \left(\frac{10^{-9}\,{\rm Hz}}{f}\right)^2 M_\odot.
\ee

The variance of  the density field $\delta(\vec x)$ is given by
\be
\sigma^2_\delta =\int_0^\infty {\rm d}\ln k \,T^2(k, R_H) W^2(k,R_H){\cal P}_{\delta}(k),
\ee
in terms of the density power spectrum ${\cal P}_{\delta}(k)$.
A real space top-hat window function $W(k,R_H)$ is introduced to smooth out the density contrast on the comoving horizon length  $R_H\sim 1/aH$, given by \footnote{
As pointed out in Ref.~\cite{sy}, the uncertainty in the choice of the window function is reduced if the same smoothing is adopted in the calculation of the threshold. For such a reason we have chosen $\delta_c=0.51$, see Tab.~I in Ref.~\cite{sy}.
}
\be
W(k,R_H) = 3 \frac{\sin (k R_H)-  (k R_H) \cos (k R_H)}{ (k R_H)^3},
\ee
 and the transfer function during radiation domination with constant degrees of freedom is provided by
\be
T(k, R_H)= 3 \frac{\sin (k R_H/\sqrt{3})-  (k R_H/\sqrt{3}) \cos (k R_H/\sqrt{3})}{ (k R_H/\sqrt{3})^3}.
\ee
Notice that, in our results, we have accounted for the ineludible non-Gaussianity inherited by the non-linear relation between the 
curvature perturbation and the density contrast  as shown in Eq.~\eqref{nonlinear}, see Refs.~\cite{DeLuca:2019qsy,Young:2019yug}.

To assess if PBHs may or not represent the dark matter in the universe, one usually introduces the PBH mass function mass $f_\text{\tiny PBH}(M_\text{\tiny PBH})$
as the fraction of PBHs with $M_\text{\tiny PBH}$  \cite{revPBH}
 \be
f_\text{\tiny PBH}(M_\text{\tiny PBH})  = \frac{1}{\Omega_{\text{\tiny DM}}} \frac{\d \Omega_\PBH}{\d \ln M_\text{\tiny PBH}}, 
\ee
such that the total fraction of dark matter in the form of PBHs is given by
\be
f_\text{\tiny PBH} =  \int f_\text{\tiny PBH}(M_\text{\tiny PBH}) \d \ln M_\text{\tiny PBH}.
\ee
After matter-radiation equality, it can be expressed in terms of the mass fraction $\beta$ as (see for example \cite{Byrnes:2018clq})
\be
f_\text{\tiny PBH}(M_\text{\tiny PBH}) = \frac{1}{\Omega_\text{\tiny DM}} \lp \frac{M_\text{\tiny eq}}{M_\text{\tiny PBH}} \rp^{1/2} \beta (M_\text{\tiny PBH}),
\ee
where the overall factor, dependent on the horizon mass at matter-radiation equality $M_\text{\tiny eq} = 2.8 \cdot 10^{17} M_\odot$,  accounts for the energy density evolution during the remaining  radiation-dominated phase after formation of PBHs of mass $M_\PBH$.

In the following sections, we will show that a signal as the one observed by NANOGrav naturally arises from a class of models with a broad 
and flat power spectrum of the curvature perturbation of the form \cite{DeLuca:2020ioi}
\be\label{A}
\Pz(k)\approx A_\zeta\, \Theta (k_s - k) \Theta (k - k_l), \qquad k_s \gg k_l
\ee
where $\Theta$ is the Heaviside step function and $A_\zeta$ is the amplitude of the power spectrum. This shape may be generated naturally for modes which exit the Hubble radius during a non-attractor phase, obtained through
an ultra slow-roll regime of the inflaton potential, as a result of a duality transformation which maps the non-attractor phase into a slow-roll phase \cite{Wands:1998yp,Leach:2000yw,Leach:2001zf, Biagetti:2018pjj}. 
 
\vskip 0.3cm
\noindent

\paragraph{The power spectrum of GWs.}
To investigate the GW signal, we first define the  linearized line element in tightly-coupled radiation domination as
\begin{equation}
 \d s^2\! =\! a^2\left\{-(1+2\Psi){\rm d}\eta^2 + \left[(1-2\Psi)\delta_{ij}+ \frac{h_{ij}}{2}\right]{\rm d}x^i {\rm d}x^j\right\},
\end{equation}
in terms of the Newtonian-gauge scalar metric perturbation $\Psi$ and the transverse-traceless tensor metric perturbation $h_{ij}$.
Focusing on the signal sourced at second-order by linear scalar perturbations \cite{Acquaviva:2002ud, Mollerach:2003nq, Ananda:2006af, Baumann:2007zm,saito, bell,Cai:2018dig,Bartolo:2018rku, Bartolo:2018evs,Unal:2018yaa,Bartolo:2019zvb,Wang:2019kaf,Cai:2019elf, DeLuca:2019ufz, Inomata:2019yww,Yuan:2019fwv,Pi:2020otn,Yuan:2020iwf}, one can expand the Einstein's equations and determine
the equation of motion for the GWs as
\begin{equation}
h_{ij}''+2\mathcal H h_{ij}'-\nabla^2 h_{ij}=-4 \mathcal T_{ij}{}^{\ell m}\mathcal S_{\ell m},
\label{eq: eom GW1}
\end{equation}
where $'$ is the derivative with respect to the  conformal time $\eta$ and
$\mathcal T_{ij}{}^{\ell m}$ projects the source term $\mathcal S_{\ell m}$ into its transverse and traceless part.
In the radiation phase the source is given by~\cite{Acquaviva:2002ud}
\begin{equation}
\label{psi}
\mathcal S_{ij}=2\partial_i\partial_j\left(\Psi^2\right)-2\partial_i\Psi\partial_j\Psi-\partial_i\left(\frac{\Psi'}{\mathcal H}+\Psi\right)\partial_j\left(\frac{\Psi'}{\mathcal H}+\Psi\right),
\end{equation}
while the projector in Fourier space using the chiral basis is 
\begin{equation}
\FF{\mathcal T}_{ij}{}^{\ell m}(\vk)=e_{ij}^\text{\tiny L}(\vk)\otimes e^{\text{\tiny L}\ell m}(\vk)
  + e_{ij}^\text{\tiny R}(\vk)\otimes e^{\text{\tiny R}\ell m}(\vk),
\end{equation}
where $e_{ij}^\text{\tiny L,R}$ are the polarisation tensors.
\begin{figure*}[t!]
	\includegraphics[width=0.48\linewidth]{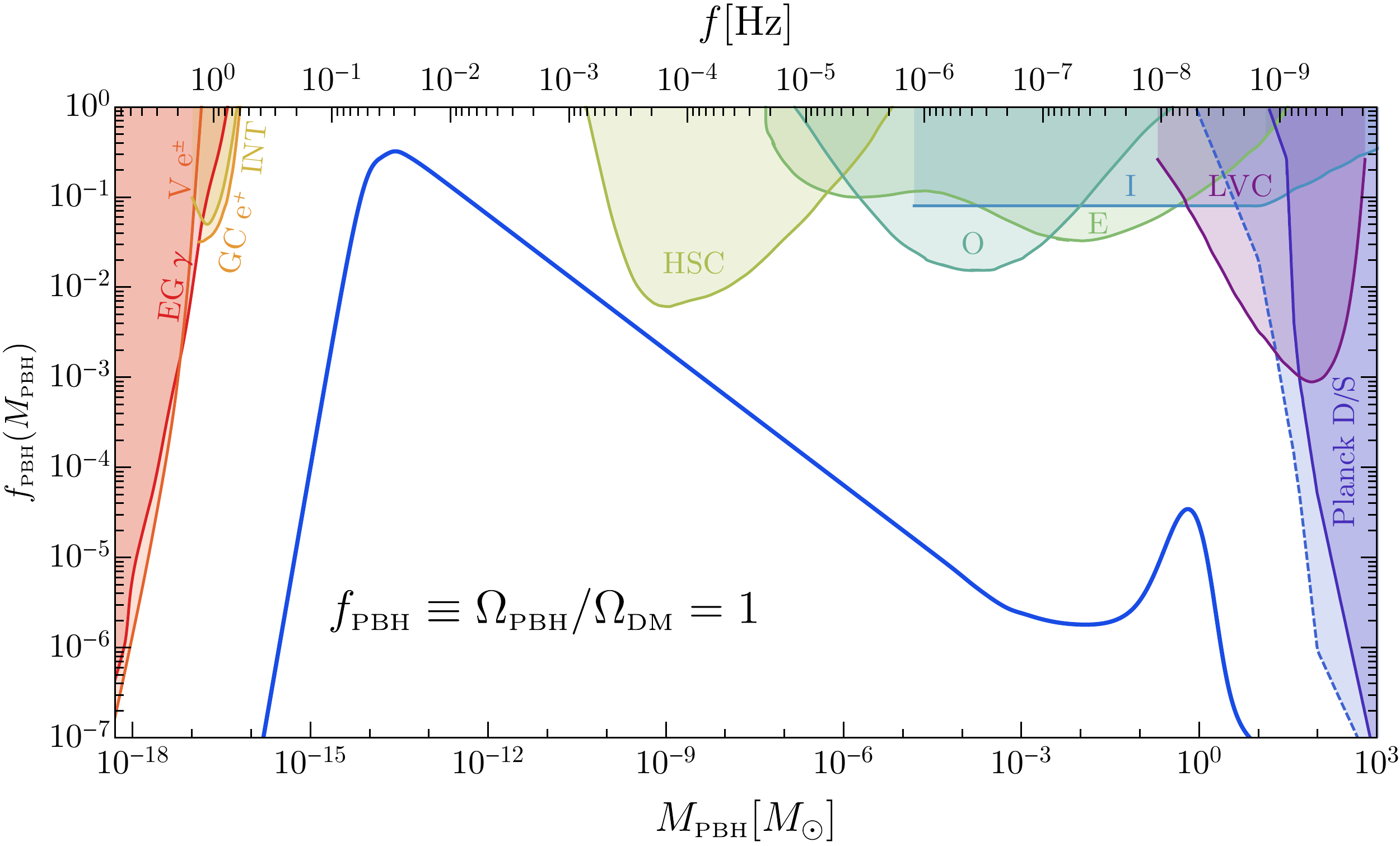}
	\includegraphics[width=0.489\linewidth]{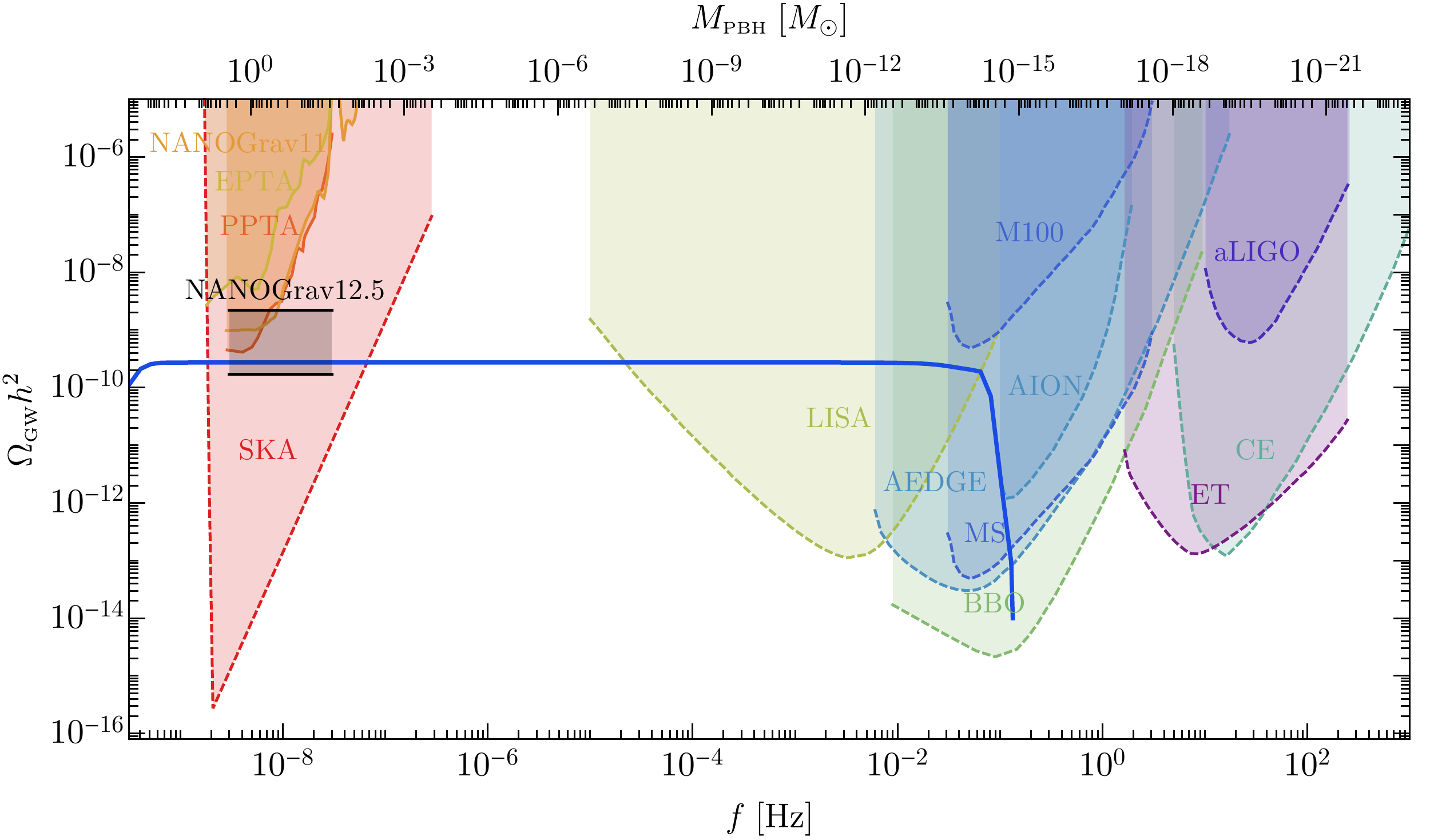}
	\caption{
\textit{{\it Left:}}
		Mass function resulting from a flat power spectrum such that it peaks at $\simeq 10^{-14} M_\odot$, with $A_\zeta \simeq 5.8\cdot 10^{-3}$ and $k_s=10^{9} k_l\simeq 1.6$ Hz, and PBHs comprise the totality of DM, i.e. $f_\PBH=1$.
	In the tail of the population, around $M_\odot$, one can notice the bump in the PBH production due to the decrease of the threshold by QCD epoch equation of state \cite{Jedamzik:1996mr,Byrnes:2018clq}.
Shown are the most stringent constraints in the mass range of phenomenological interest coming from the Hawking evaporation producing
extra-galactic gamma-ray (EG $\gamma$) \cite{Arbey:2019vqx}, $e^\pm$ observations by Voyager 1 (V $e^\pm$) \cite{Boudaud:2018hqb}, positron annihilations in the Galactic Center (GC $e^+$) \cite{DeRocco:2019fjq} and gamma-ray observations by INTEGRAL (INT) \cite{Laha:2020ivk} (for other constraints in this mass range see also \cite{Carr:2009jm, Ballesteros:2019exr,Laha:2019ssq, Poulter:2019ooo,Dasgupta:2019cae,Laha:2020vhg}), microlensing searches by Subaru HSC \cite{Niikura:2017zjd, Smyth:2019whb}, MACHO/EROS \cite{Alcock:2000kd, Allsman:2000kg}, Ogle \cite{Niikura:2019kqi} and Icarus \cite{Oguri:2017ock}, and those coming from CMB distortions by spherical or disk accretion (Planck S and Planck D, respectively) \cite{Ali-Haimoud:2016mbv, Serpico:2020ehh}. 
LVC stands for the constraint coming from LIGO/Virgo Collaboration measurements \cite{Ali-Haimoud:2017rtz,raid,ver}. We neglect the role of accretion which has been shown to affect constraints on masses larger than ${\cal O}(10) M_\odot$ \cite{c1,c2}.
See Ref.~\cite{Carr:2020gox} for a comprehensive review on constraints on the PBH abundance. Notice that there are no stringent constraints in the PBH mass range of interest \cite{Katz:2018zrn, Montero-Camacho:2019jte}.
\textit{{\it Right:}}
The abundance of GWs according to our scenario. In black the 95\% C.I. from the NANOGrav 12.5 yrs experiment is shown. For more details about the projected sensitivities see the main text.
}
	\label{figs}
\end{figure*}
In Eq. (\ref{psi}) the scalar perturbation $\Psi(\eta,\vk)$ can be expressed in terms of the comoving curvature perturbation as~\cite{lrreview}
\begin{equation}
\Psi(\eta,\vk) \equiv \frac 23 T(k\eta) \zeta(\vk) .
\label{eq: Psi to zeta}
\end{equation}
The current abundance of GWs is found to be \cite{errgw}
\begin{eqnarray}
\frac{\OGW(f)}{\Omega_{r,0}} &= &\frac{c_g}{972}
  \int \int_\mathcal{S} \d x \d y
  \frac{x^2}{y^2} \left[1 - \frac{(1+x^2-y^2)^2}{4x^2} \right]^2 \nonumber \\
& \times & \Pz \left(k x \right) \Pz\left(k y\right){\cal I}^2(x,y),
\label{eq: Omega GW with PS0}
\end{eqnarray}
where the integration region $\mathcal{S}$ extends to $x > 0$ and to $|1 - x | \leq y  \leq 1 +x$, and $k = 2 \pi f$. 
The parameter $c_g$ defined as 
\begin{equation}
c_g	\equiv \frac{g_*(M_H)}{g_{*}^0}\lp \frac{g_{*S}^0}{g_{*S} (M_H)}\rp ^{4/3}
\end{equation}  accounts for the change of the effective degrees of freedom of the thermal radiation $g_*$ and $g_{*S}$ (where the superscript $^0$ indicates the values today) during the evolution (assuming Standard Model physics), and it is of order $c_g=0.4$ for modes related to the formation of asteroid-mass PBHs.
Also, $\Omega_{r,0}$ stands for the current  radiation density if the neutrinos were massless, ${\cal I}^2\equiv\Ic^2+\Is^2$ and
\begin{align}
& \Ic(x,y) =4  \int_0^\infty \dd\tau \, \tau (-\sin \tau)  \Big[ 2T(x\tau)T(y\tau) \nonumber\\
&+ \Big(T(x\tau)
+ x\tau\, T'(x\tau) \Big)\Big(T(y\tau) + y\tau\, T'(y\tau) \Big) \Big],
\label{eq: Ic, Is}
\end{align}
$\Is(x,y)$  being the same function, but with $(-\sin \tau)$ replaced by $(\cos\tau)$, see Refs.~\cite{errgw,Kohri:2018awv}.

\noindent
\vskip 0.3cm
\paragraph{Results.}
We have collected our results in Fig.~\ref{figs}. In the left panel,  we have plotted the mass function corresponding to the primordial curvature perturbation given in Eq.~\eqref{A}. As described in Ref.~\cite{DeLuca:2020ioi}, the peak of the mass function for a broad flat spectrum \eqref{A} corresponds to the mass inside the horizon when the shortest scale $\sim 1/k_s$  re-enters the horizon. At smaller masses, the mass function goes as $M_\PBH ^{3.8}$ due to the dinamics of the critical collapse, while at larger masses  falls down as $\sim M_\PBH^{-3/2}$ and has a sub-dominant peak around $\sim M_\odot$ due to the change of equation of state during the QCD phase transition \cite{Jedamzik:1996mr,Byrnes:2018clq}. Given the absence of constraints in the mass range of support of the PBH mass function
( the femtolensing bounds have been shown to be inconsistent once the extended nature of the source as well as wave optics effects are properly taken into account \cite{Katz:2018zrn, Montero-Camacho:2019jte}), the integral of the latter  can be chosen in such a way that the PBHs contribute to the totality of the dark matter, that is
\begin{equation}
 f_\text{\tiny PBH}=\int f_\text{\tiny PBH}(M_\text{\tiny PBH}) \d \ln M_\text{\tiny PBH}=1.
\end{equation}
As a consequence the first prediction of our scenario is that the signal seen by NANOGrav, if interpreted as a stochastic background of GWs produced as second-order within the PBH model, is in agreement with the possibility that all the dark matter is in the form of extremely light PBHs.

On the right panel of Fig.~\ref{figs}, we show the corresponding spectrum of the second-order GW abundance as a function of the frequency which falls within the 
95\% C.I. from the NANOGrav 12.5 yrs observation.
Shown are the constraints coming from experiment EPTA \cite{Lentati:2015qwp}, PPTA \cite{Shannon:2015ect}, NANOGrav 11 yrs \cite{Arzoumanian:2018saf, Aggarwal:2018mgp} and future sensitivity curves for planned experiments like SKA \cite{ska}, LISA \cite{Audley:2017drz} (power-law integrated sensitivity curve expected to fall in between the designs named C1 and C2 in Ref.~\cite{caprini}), DECIGO/BBO \cite{BBO}, CE \cite{Evans:2016mbw}, Einstein Telescope \cite{ET-1, ET-2}, Advanced Ligo + Virgo collaboration \cite{ligo}, Magis-space and Magis-100 \cite{magis-100}, AEDGE \cite{aedge} and AION \cite{aion}.
Notice that a portion of the 95\% C.I. of NANOGrav 12.5 yrs is in tension with NANOGrav 11 yrs and PPTA. However, according to the NANOGrav Collaboration \cite{Arzoumanian:2020vkk} the improved priors for the intrinsic pulsar red noise used in the novel analysis relaxes the NANOGrav 11 yrs bound. Nevertheless, the predicted signal within our scenario falls below all bounds. The GW abundance spectrum propagates flat entering the LISA detectable region and decays rapidly at the frequency corresponding to the shortest scale $1/k_s$. The second prediction of our scenario is therefore that the second-order GWs seen by NANOGrav should also be detected by the forthcoming experiment LISA, and eventually MS and BBO as well.
Notice also that the present scenario is consistent with the candidate event found by the HSC collaboration and discussed in Ref.~\cite{kus}.

Both predictions of the scenario described in this paper depend only on the choice of the shortest scale $1/k_s$ and the requirement of the PBH abundance being equal to the dark matter one.

\noindent
\vskip 0.3cm
\paragraph{Conclusions.} 
The discovery of a primordial stochastic background of GWs would be another fundamental pillar in GW astronomy. In this Letter, we have shown that the recently published  stochastic common-spectrum process by the NANOGrav Collaboration, if interpreted as an indication of a GW background, can be naturally linked to the physics of PBHs. Indeed, the formation of PBHs in the early universe due to the collapse of sizeable overdensities generated during inflation is inevitably accompanied by the generation of GWs. Interestingly enough, the NANOGrav observation is consistent with a mass range of PBHs such that the latter can comprise the totality of the dark matter. Furthermore, the GW signal is characterized by a flat spectrum which will make it visibile even at much larger frequencies, most notably by the forthcoming experiment LISA.

\noindent
\vskip 0.3cm
\paragraph{Note Added.}
When completing this work, other interpretations of the NANOGrav signal have appeared other than  through super-massive BH coalescences \cite{ses}. Refs. \cite{Blasi:2020mfx,Ellis:2020ena} point out the possible generation of GWs through
 cosmic strings. 
Ref.~\cite{vv}  discusses as well the possibility of explaining the NANOGrav signal with second-order GWs related to the PBH scenario. They identify  a  mass range of PBHs providing  the seeds of supermassive BHs which cannot be the dark matter. Contrarily to that work, in our scenario the PBH masses are much smaller and PBHs can comprise the totality of the dark matter.
The possible confirmation of the scenario proposed in this work by future HSC observations has been forcasted in Ref.~\cite{Sugiyama:2020roc}.

\vskip 0.3cm
\noindent
\paragraph{Acknowledgments.}
\noindent
 We thank Ranjan Laha, Marek Lewicki and Julian Mu\~noz for discussions.
V.DL., G.F. and 
A.R. are supported by the Swiss National Science Foundation 
(SNSF), project {\sl The Non-Gaussian Universe and Cosmological Symmetries}, project number: 200020-178787.

\bigskip


\end{document}